\documentclass[pre,aps,onecolumn]{revtex4}

\usepackage{graphicx}
\usepackage{amsmath}
\usepackage{amssymb}
\usepackage{ifthen}
\usepackage{booktabs}
\usepackage{color}

\usepackage{graphicx}
\usepackage{dcolumn}
\usepackage{bm}
\usepackage{longtable}
\usepackage{gensymb}

\newcommand{\bb}{\begin{equation}}
\newcommand{\ee}{\end{equation}}
\newcommand{\ba}{\begin{eqnarray*}}
\newcommand{\ea}{\end{eqnarray*}}
\newcommand{\rhor}{\rho({\bf r})}
\newcommand{\dd}{{\rm d}}
\newcommand{\rr}{{\mathbf r}}
\newcommand{\dr}{{\rm d}{\bf r}}

\usepackage{graphicx}
\usepackage{dcolumn}
\usepackage{bm}
\usepackage{longtable}

\begin{document}


\title{Effective interactions between a pair of nanoparticles}

\author{Alexandr \surname{Malijevsk\'y}}
\affiliation{ {Department of Physical Chemistry, Institute of Chemical Technology, Prague, 166 28 Praha 6, Czech Republic}}

\begin{abstract}
We investigate the effective interactions between two nanoparticles (or colloids) immersed in a solvent exhibiting two-phase separation. Using a non-local density
functional theory, we determine the dependence of the effective potential on the separation of the nanoparticles when the solvent is near bulk two-phase coexistence.
If identical nanoparticles preferentially adsorbing phase $\alpha$ are inserted into phase $\beta$, thick wetting layers of the preferable phase $\alpha$ develop at
their surfaces. At some particular separation $h_b$ of the nanoparticles, the wetting layers connect to form a single bridge, and the induced effective potential
becomes strongly attractive for all distances $h<h_b$. The bridging is a first order capillary condensation like transition for all radii of the nanoparticles greater
than the critical radius $R_c$, the value of which was estimated to be approximately $R_c\approx20\sigma$ for a temperature $T/T_c\approx0.9$, where $\sigma$ is the
size of the solvent (square-well) particles. For radii $R<R_c$ the process of bridging is continuous. If the same particles are inserted into the preferable phase
$\alpha$, the only effective interaction between them is induced by the short-ranged depletion potential. If the nanoparticles have opposite adsorption preferences,
only a single wetting layer forms around one of the nanoparticles and the effective interaction is strongly repulsive in both phases. The repulsion, induced by a
disruption of the wetting film by the presence of the second particle, is larger and slightly longer-ranged in a low density state.

\end{abstract}


\maketitle

\section{Introduction}

The study of nanoparticles (or colloids) dispersed in a sea of  solvent particles  is a popular topic of research in the statistical physics of liquids. One reason
for this popularity is the possibility of applying coarse-graining techniques integrating out certain degrees of freedom of the complex system, which may obey a
relatively simple interaction model that, on one hand, still possesses most of the crucial physics and, on the other, is computationally tractable via well developed
methods from the statistical theory of simple fluids. A well-known example presents the simple model introduced originally by Asakura and Oosawa for a mixture of
colloids and non-adsorbing polymers \cite{ao1,ao2, vrij}. This model represents an extreme case of a non-additive hard-sphere mixture where the colloids are treated
as hard spheres and the polymers are modelled as point particles that are excluded from colloids by a centre-of-mass distance reflecting the colloid diameter and the
radius of gyration of the polymer coils. Another important model is a binary mixture of highly size-asymmetric hard spheres, such that $q=\sigma_b/\sigma\gtrsim10$
where $\sigma_b$ and $\sigma$ are the diameters of the large and the small species, respectively. Here, the smaller species can again represent non-adsorbing polymers
or a second species of the colloidal mixture. In both cases, the depletion potential arises between a pair of large particles due to excluding volume effects that
give rise to the osmotic pressure between the bulk region of the small species and the depleted region between two adjacent large spheres \cite{lek1}. For the
Asakura-Oosawa model, the depletion potential is attractive within a short distance $r<\sigma_b+\sigma$ between the centres of the colloidal particles and vanishes
beyond this separation. For a binary mixture of hard spheres, the additional excluded volume effects between the small particles may significantly change this
scenario especially at a high concentration of the small spheres and as a result, the depletion potential has a more complicated oscillatory structure, especially at
high densities. In any case, the depletion potential embodies an entropically driven interaction between large particles complementing the bare hard-sphere potential.
This extra effective attraction plays an important role for phase separation \cite{lek, biben, dijk} and flocculation \cite{jenkins, somas} in colloidal and
colloid-polymer mixtures.

Another type of solvent mediated interactions between two large bodies (nanoparticles) can occur in solvents exhibiting fluid coexistence. Depending on the
thermodynamic conditions, wetting layers of a given thickness may adsorb at the nanoparticle surfaces, giving rise to a solvent-mediated potential, quite distinct
from the depletion potential. First of all, because the wetting layers can be fairly thick at or near the solvent two phase coexistence and for large radii of the
solute particles, the range of this interaction can be substantially longer than that of the depletion potential. Furthermore, in view of the much higher number of
solvent particles contributing to the interaction, the resulting net force between the nanoparticles is expected to be much stronger than that of the depletion force.
This, indeed, has been corroborated in relatively recent theoretical studies \cite{archer, hopkins} for soft-core models. Finally, and perhaps most importantly, upon
drawing the nanoparticles coated by their wetting layers closer to each other, the bridging transition occurs as the wetting layers merge into a single film
encapsulating both nanoparticles. The transition is generally deemed to be first order (at least, on a mean-field level) and leads to a strongly attractive bridged
configuration. Here, it should be noted that by the term ``wetting layer'' we do not necessarily mean a liquid layer; it can be either a liquid or a gas phase
intruding between a nanoparticle and a bulk solvent if the solvent exhibits a liquid-gas coexistence or it can be a preferentially adsorbed phase if the solvent
exhibits a fluid-fluid separation.

 In this work we investigate the effective interactions between two spherical nanoparticles dispersed in a one-component fluid (solvent) exhibiting liquid-vapour
coexistence using a microscopic fundamental measure density functional theory (FM-DFT). The solvent particles are modelled by a square-well (SW) fluid that is treated
grand canonically.  The pair of nanoparticles is at a fixed mutual separation and acts as an external field, which leads to a 2D axially-symmetric problem. Attempts
to describe the model as a 1D problem led to a formulation of the so-called insertion method \cite{roth}, in which case the use of the exact potential distribution
theorem  allows to treat the system as a binary mixture in an external field exerted by only a single nanoparticle. Although this approach simplifies the numerics of
the problem significantly and provides a good approximation for the effective interaction for the state points far from the coexistence, the insertion method was
found inadequate to account for the effects of bridging \cite{archer}. For the solvent models interacting via dispersion forces, the use of the sharp-kink
approximation was made to determine the solvent mediated interaction between the nanoparticles \cite{yeomans, yeomans2, bauer}.

Within Model 1, the nanoparticles are represented by two identical hard-sphere (HS) particles of sizes much larger than that of the solvent SW fluid. Model 2 mimics
the situation where the two nanoparticles are competing in the sense that different phases are preferably adsorbed at their surfaces. In this case, one of the
nanoparticles is again a hard sphere but the other one is the square-well particle, such that their hard cores are of identical diameters (and again much larger than
that of the solvent particles). We fix a subcritical temperature of the bulk solvent and immerse the nanoparticles into either of the coexisting fluid phases of the
solvent. Within Model 1 and the liquid solvent, the HS particles, when sufficiently far apart, experience complete drying (wetting by gas). The finite curvature of
the nanoparticles prevents the drying films from divergening even at the two-phase coexistence, as it is well known that the wetting layer thickness grows
logarithmically with (large) $R$ for systems with short-range potentials \cite{holyst}. The situation here is similar to the one studied in Ref. \cite{hopkins} for a
binary mixture of soft-core solvent particles exhibiting fluid-fluid separation and treated by a simple DFT where a mean-field approximation was used  for the entire
part of the excess free energy functional (which corresponds to the random phase approximation for the direct correlation function). Upon bringing the large particles
closer together, the authors found a first-order transition associated with the connection of the wetting into a single bridge. This abrupt transition exhibits itself
as a cusp in the effective potential $W$ as a function of the distance between the large particles. When the bridge is formed, $W$ decreases steeply with the
distance, indicating a very strong attraction between the large particles in this regime. If, on the other hand, nanoparticles of the same type (Model 1) are inserted
into the gas phase of the bulk solvent then, of course, no wetting layers form at their surfaces. Therefore, one does not expect any substantial effective interaction
between the nanoparticles except, perhaps, for very small separations of the nanoparticles in which case depletion effects presumably play some role.

For Model 2, the situation is clearly quite different. Here, in either case, a wetting film forms at one of the nanoparticle surfaces. That is, in the case of the
liquid ambient phase, a single drying film forms around the HS particle whilst if the bulk phase is gas, the situation is just reversed with a wetting layer formed at
the SW nanoparticle. In both cases, the asymmetry in adsorption preferences must be reflected as the two nanoparticles are brought close together.

The paper proceeds as follows. In section II, we set the molecular model and formulate the FM-DFT in terms of weighted densities expressed as two-dimensional
integrals in cylindrical coordinates. In section III, we present our numerical results for the effective potentials  of both Model 1 and Model 2. In particular, we
determine a critical value for the hard-sphere diameter $R_c$ below which the first-order bridging transition disappears. We also show that for Model 2 the effective
interaction between the nanoparticles is harshly repulsive for small separations and interpret our results by using simple arguments based on the interfacial
Hamiltonian approach. In section IV, we summarise and discuss our main results and conclude by outlining a perspective for future work.

\section{Density functional theory}

Within classical density functional theory the equilibrium density profile is found by minimising the grand potential functional
 \bb
 \Omega[\rho]={\cal{F}}[\rho]+\int\dr\rhor\left[V(\rr)-\mu\right]\,.\label{om}
 \ee

where $V(\rr)$ is the external field and $\mu$ is the chemical potential. The intrinsic free energy functional ${\cal{F}}$ characterises a given model fluid
regardless of $V(\rr)$, and it is convenient to split this term into the contribution due to the ideal gas and the remaining excess part
 \bb
 {\cal{F}}[\rho]={\cal{F}}_{\rm id}[\rho]+{\cal{F}}_{\rm ex}[\rho]\,,
 \ee
where ${\cal{F}}_{\rm id}[\rho]= \beta^{-1}\int\dr\rhor\left[\ln(\Lambda^3\rhor)-1\right]$ with $\beta=1/k_BT$ and $\Lambda$ being the thermal de Broglie wavelength
that can be set to unity.

In the spirit of the van der Waals theory, the excess term is treated in a perturbative manner, such that we separate the repulsive and attractive contributions:
 \bb
{\cal{F}}_{\rm ex}[\rho]={\cal{F}}_{\rm hs}[\rho]+\frac{1}{2}\int\dr\rhor\int\dr'\rho(\rr')u(|\rr-\rr')\,.\label{fex}
 \ee
Here, the attractive portion of the excess free energy functional is treated in a mean-field fashion, while the repulsive bit is approximated by the excess HS free
energy using the fundamental measure theory \cite{ros}
 \bb
 \beta{\cal{F}}_{\rm hs}[\rho]=\int\dr\Phi(\{n_\alpha\})\,.\label{fhs}
 \ee

We consider two models characterised by their external potentials. Within Model 1, the external potential $V_1(\rr)$ is exerted by two large hard spheres, each of a
diameter $\sigma_b$,  placed a distance $\sigma_b+h$ apart. Owing to the axial symmetry of the model, the potential is expressed in the cylindrical coordinates $(z,
r)$:
 \bb
 V_1(z,r)=V_{\rm hs}(z-\sigma_b/2-h/2,r)+V_{\rm hs}(z+\sigma_b/2+h/2,r)\,,\label{v1}
 \ee
 where
  \bb
 V_{\rm hs}(z,r)=\left\{\begin{array}{cc}
 \infty& z^2+r^2\leq R^2\,,\\
 0\,;&z^2+r^2>R^2\,,
\end{array}\right.
 \ee
 and $R=\sigma_b/2$.

In Model 2, one of the HS particles has an additional SW attraction, such that the entire external potential is:
 \bb
 V_2(z,r)=V_{\rm hs}(z-\sigma_b/2-h/2,r)+V_{\rm sw}(z+\sigma_b/2+h/2,r)\,,\label{v2}
 \ee
 where
  \bb
 V_{\rm sw}(z,r)=\left\{\begin{array}{cc}
 \infty& z^2+r^2\leq R^2\,,\\
  -\varepsilon\,;&R^2<z^2+r^2<(R+\sigma)^2\,.\\
 0\,;&z^2+r^2\geq(R+\sigma)^2\,.
\end{array}\right.
 \ee
is expressed in terms of the hard core diameter $\sigma$ and the depth of the well $\varepsilon$ of the solvent SW fluid. The attractive portion of the fluid-fluid
interaction that enters  Eq.~(\ref{fex}) is:
  \bb
 u(\tilde{r})=\left\{\begin{array}{cc}
 -\varepsilon& \tilde{r}<\lambda\sigma\,,\\
 0\,;&\tilde{r}\geq\lambda\sigma\,,
\end{array}\right.
 \ee
where $\tilde{r}=\sqrt{(z_2-z_1)^2+(r_2-r_1)^2}$ is the distance between the centres of two SW particles and $\lambda=1.5$, as usual.

 The repulsive part of the intrinsic free energy given by Eq.~(\ref{fhs}) corresponds to a one-component HS fluid of diameter $\sigma$. The free energy density
$\Phi$ is a function of a set of weighted densities $\{n_\alpha\}$, five of which are given by surface integrals that must be expressed as two-dimensional integrals.
The remaining weighted density $n_3$ is given by integrating the density profile $\rho(z, r)$ over the volume of a hard core of diameter $\sigma$. To minimise the
dimensionality of this volume integral by exploiting the axial symmetry of the system, we can express $n_3$ as a double integral as follows:

 \bb n_3(z,r)=2\int_{-R}^R\dd z' \int_{r-\tilde{R}(z')}^{r+\tilde{R}(z')}\dd r' r'
\arccos\left[\frac{r'^2+r^2-\tilde{R}^2(z')}{2r'r}\right]\rho(z'+z,r');\;\;\;r>R \label{n3a}
 \ee

\bb
 n_3(z,r)=2\int_{-R}^{R}\dd z' \Psi(z';z,r)+2\pi\int_{-\sqrt{R^2-r^2}}^{\sqrt{R^2-r^2}}\dd z'\int_0^{\tilde{R}(z')-r}\dd r'r'\rho(z'+z,r')\nonumber;\;\;\;r<R\,, \label{n3b}
\ee
 with
 \bb
\Psi(z';r,z)\equiv\int_{|r-\tilde{R}(z')|}^{r+\tilde{R}(z')}\dd r' r' \arccos\left[\frac{r'^2+r^2-\tilde{R}^2(z')}{2r'r}\right]\rho(z'+z,r')\,.
 \ee
 and where $\tilde{R}(z)=\sqrt{R^2-z^2}$.

The minimisation of Eq.~(\ref{om}) leads to the Euler-Lagrange equation
 \bb
 \frac{1}{\beta}\ln\Lambda^3\rho(z,r)+\frac{\delta F_{\rm ex}}{\delta\rho(z,r)}+V(z,r)=\mu \label{el}
 \ee
which is solved iteratively on a two-dimensional grid  of  a cylindrical system of  volume $V=2L_z\pi L_r^2$, where $L_r=20\sigma$ and $L_z=40\sigma$, with a uniform
spacing of $0.02\sigma$ in both $z$ and $r$ directions. Although the grid does not follow the symmetry of the external field, the fine spacing avoids any noticeable
artifacts in the structure of the fluid near the nanoparticles. By determining the equilibrium density profile, the excess (over ideal) grand potential $\Omega_{\rm
ex}$ can be calculated. The effective potential between the pair of immersed particles a distance $h$ apart is given by
 \bb
 W(h)=\Omega_{\rm ex}(h)-2\Omega_{\rm ex}^{(\rm hs)}
 \ee
 for Model 1 and by
  \bb
 W(h)=\Omega_{\rm ex}(h)-\Omega_{\rm ex}^{(\rm hs)}-\Omega_{\rm ex}^{(\rm sw)}
 \ee
 for Model 2. Here, we defined $\Omega_{\rm ex}^{\rm hs}$ and $\Omega_{\rm ex}^{\rm sw}$ as the excess grand potentials corresponding to a system with only one
 HS or one large SW particle, respectively.

\section{Results}

Throughout this work, we consider a subcritical temperature $T/T_c=0.89$ where $k_BT_c\approx0.9\varepsilon$ is the critical temperature of the bulk SW fluid, as
determined from Eq.~({\ref{el}) in the case of zero external potential. We begin with Model 1, such that two hard spheres of diameter $\sigma_b=10\,\sigma$ are
immersed in a saturated gas of the square-well fluid. In this case, no wetting/drying films develop around the HS bodies, so that the effective potential displayed in
Fig.~1 as a function of the separation between the hard-sphere surfaces is very short-range and attractive due to the depletion effect. In view of the low-density
state of the ambient fluid, the depletion potential is weak and monotonic.

\begin{figure}
\includegraphics{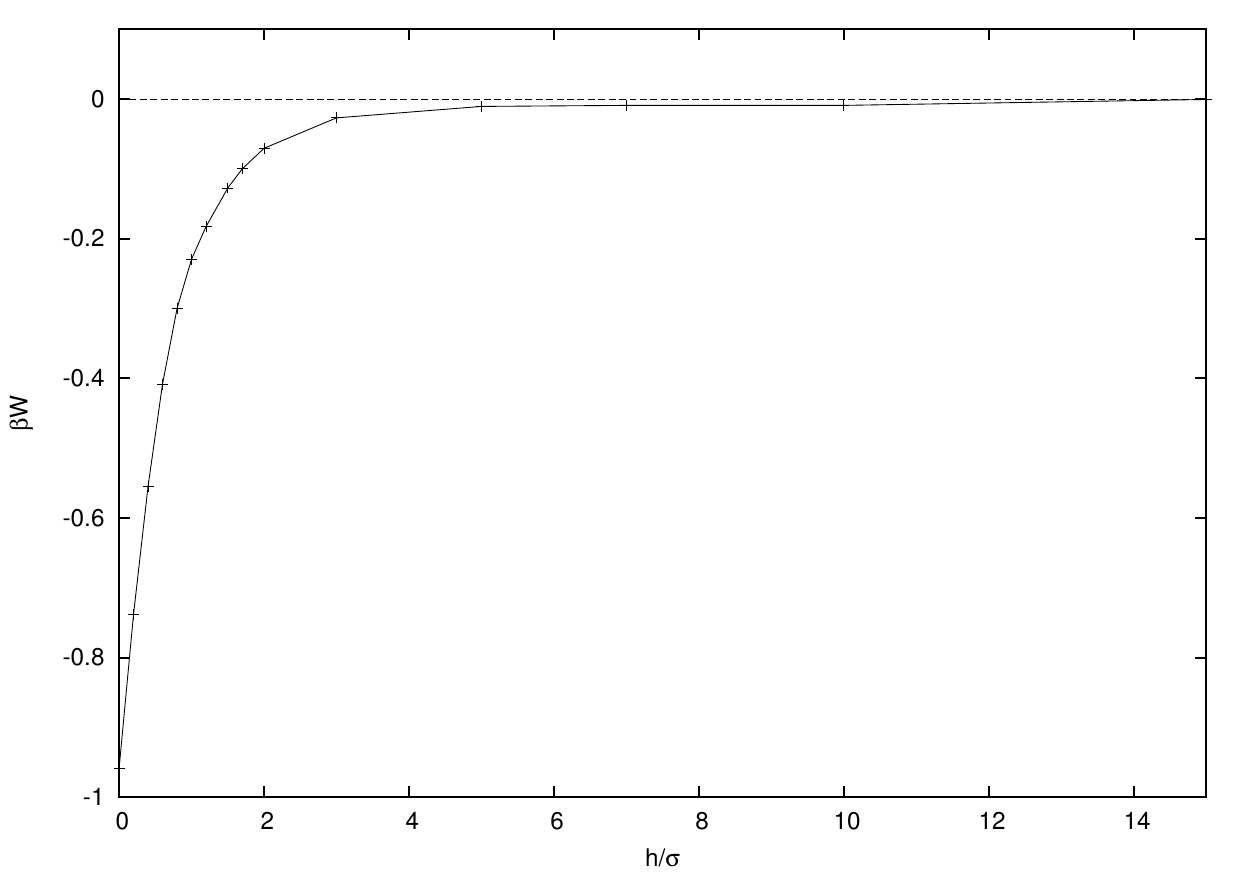}
\caption{Effective potential of a pair of hard spheres of diameters $\sigma_b=10\,\sigma$ as a function of the distance between their surfaces. The pair is immersed
in a saturated vapour of the SW fluid at a temperature $T/T_c=0.89$. } \label{fig1}
\end{figure}

\begin{figure}
\includegraphics[width=0.5\textwidth]{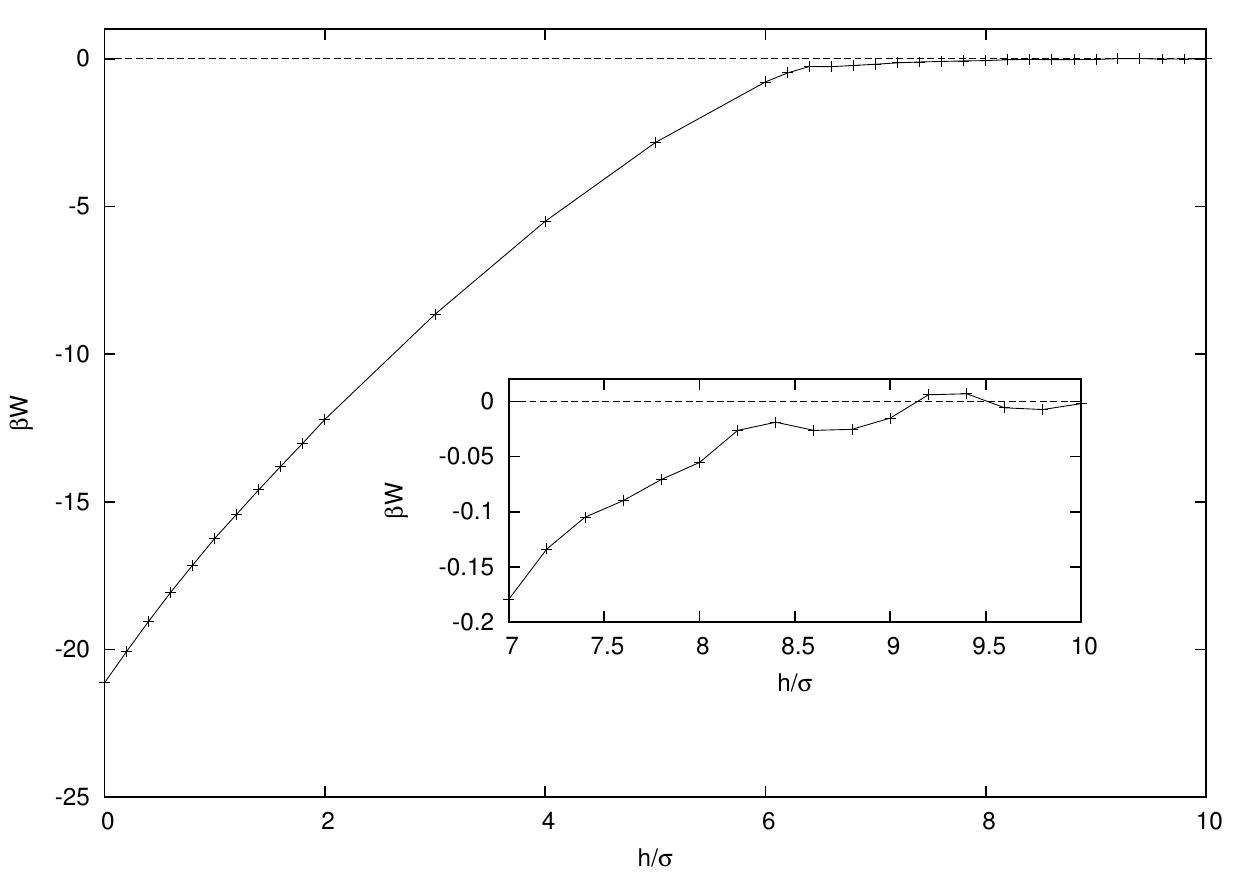}\includegraphics[width=0.5\textwidth]{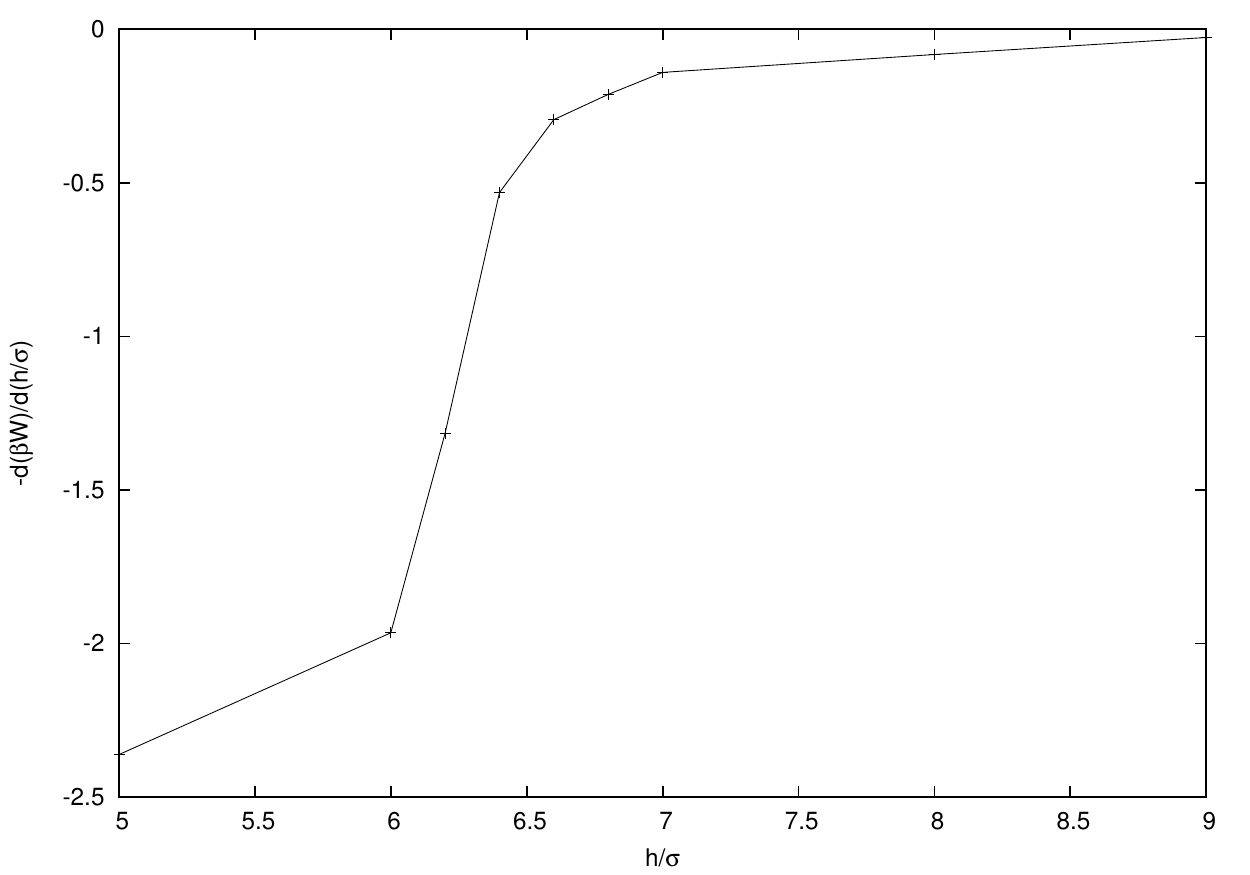}
 \caption{ Left panel: Effective potential of a pair of hard spheres of diameters $\sigma_b=10\,\sigma$ as a function of the distance between their surfaces. The pair is immersed
in a saturated liquid of the SW fluid at a temperature $T/T_c=0.89$.  The inset shows an oscillatory behaviour of the potential at larger distances. Right panel:
Effective force $-{dW}/{dh}$ between the nanoparticles. As the bridge between the nanoparticles is formed the force changes abruptly but still continuously.}
\label{fig2}
\end{figure}

\begin{figure}
\includegraphics[width=0.4\textwidth]{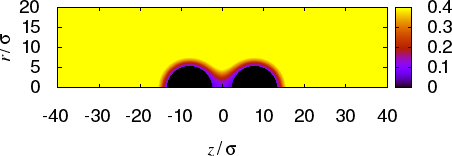}\hspace*{0.2cm}\includegraphics[width=0.4\textwidth]{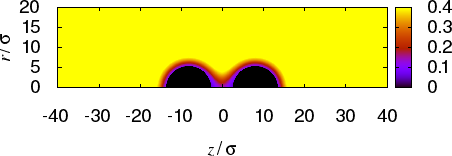}
\includegraphics[width=0.4\textwidth]{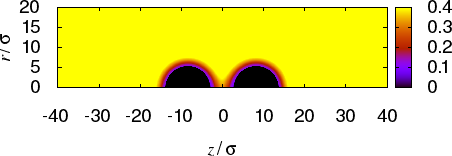}\hspace*{0.2cm}\includegraphics[width=0.4\textwidth]{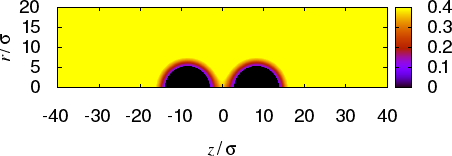}
\caption{Two-dimensional density profiles of a system consisting of two hard spheres of diameters $\sigma_b=10\,\sigma$ immersed in a saturated liquid of the SW fluid
at a temperature $T/T_c=0.89$. The separations between the hard spheres are (from top left to right bottom): $h=6\,\sigma$, $h=6.4\,\sigma$, $h=6.8\,\sigma$ and
$h=7\,\sigma$. A low density gas film develops around the spheres and the films get connected at lower separations to form a bridge. As the separation between the
hard spheres increases, the bridging bond disappears continuously.} \label{fig3}
\end{figure}

Considerably stronger and longer-ranged interactions are induced when the hard spheres are inserted into a disfavoured liquid phase. Here, we have considered four
different hard-sphere diameters $\sigma_b/\sigma=5,10,15$ and $20$. Qualitatively similar results have been obtained for the diameters $\sigma_b/\sigma\leq15$, as
illustrated in Fig.~2 for $\sigma_b=10\,\sigma$. Here, beyond a certain distance $h_b\approx6\sigma$, the effective potential is only very slightly negative but below
$h_b$ the potential rapidly decreases with the distance. Concomitant with the development of the strong interaction between the hard spheres is a formation of a
bridging bond between the particles  at a distance of approximately $h_b$. Although the curvature of the function $W(h)$ changes  abruptly near $h_b$, the function
itself is smooth and does not exhibit a hysteresis. This has been checked by performing two sets of calculations with bridged and unbridged initial configurations
that both resulted in identical minima of the grand potential functional.  In Fig.~2 we also display the effective force $-dW/dh$ as a function of $h$ that changes
abruptly yet continuously near $h_b$. The reversible process of bridging/unbridging is illustrated in Fig.~3, where the development/termination of the bridge bond is
shown for the separations between the hard spheres near $h_b$.

\begin{figure}
\includegraphics{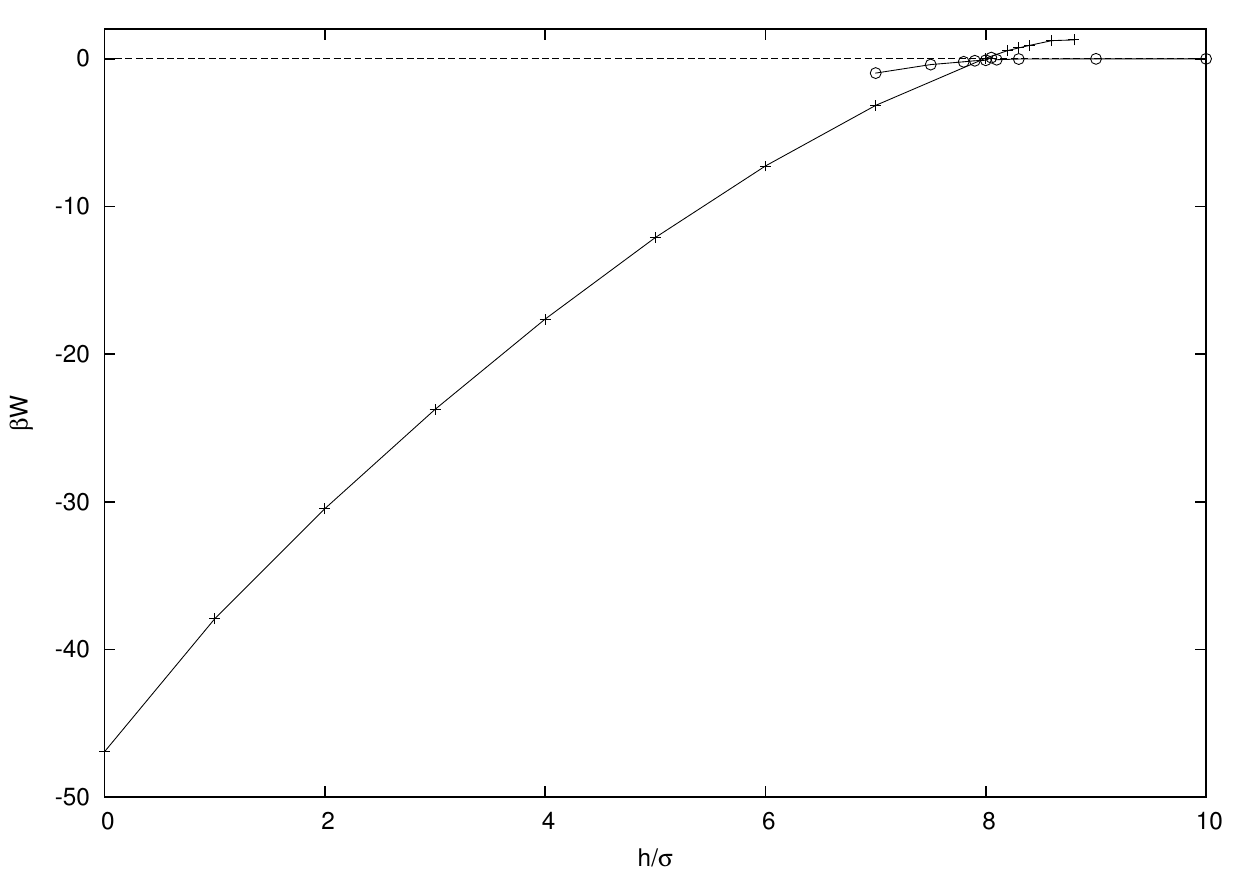}
\caption{Effective potential of a pair of hard spheres of diameters $\sigma_b=20\,\sigma$ as a function of the distance between their surfaces. The pair is immersed
in a saturated liquid of the SW fluid at a temperature $T/T_c=0.89$. The line with circles denotes the unbridged states, and the line with crosses denotes the bridged
states. The first-order bridging transition occurs at a distance $h\approx8\,\sigma$.} \label{fig4}
\end{figure}

\begin{figure}
\includegraphics[width=0.4\textwidth]{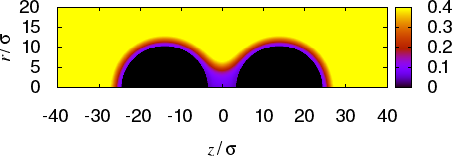}
\includegraphics[width=0.4\textwidth]{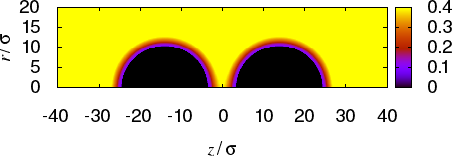} 
\caption{Two-dimensional density profiles of a system consisting of two hard spheres of diameter $\sigma_b=20\,\sigma$ immersed in a saturated liquid of the SW fluid
at a temperature $T/T_c=0.89$. In both cases, the separation between the hard spheres is $h_b=8\,\sigma$,  which corresponds to the equilibrium between the bridged
(left panel) and unbridged (right panel) configurations. } \label{fig5}
\end{figure}

The situation, however, becomes different for larger solute/solvent size ratios. In Fig.~4, we display the separation dependence of the effective potential between
two hard spheres of diameters $\sigma_b=20\,\sigma$. Starting from two different (bridged and unbridged) configurations, the minimisation of the grand potential
functional converges to two different density profiles in the vicinity of the separation $h\approx8\,\sigma$. Associated with this hysteresis are the metastable
extensions of both configurations terminating at their spinodals, such that the equilibrium state corresponds to the global minimum of the grand potential. The
intersection of these bridged and unbridged branches produces a cusp at $h_b$ corresponding to a location of a first-order bridging transition. The coexisting bridged
and unbridged configurations are shown in Fig.~5.  It should be noted that the bridge does {\emph not} form as the two wetting layers meet. Instead, in pure analogy
to a first-order capillary condensation with the walls covered by wetting films, the transition occurs at a distance $h>2\ell$ where $\ell$ is the thickness of the
wetting layer. This contrasts with the continuous process of bridging displayed in Fig.~3.

\begin{figure}
\includegraphics[width=0.22\textwidth]{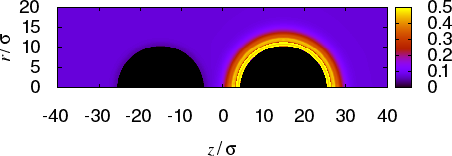}\hspace*{0.1cm}\includegraphics[width=0.22\textwidth]{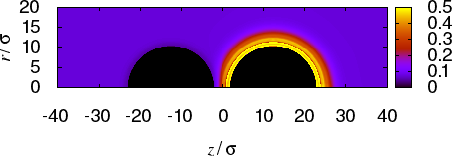}\hspace*{0.1cm}\includegraphics[width=0.22\textwidth]{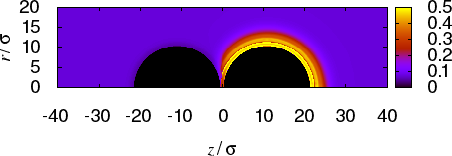}\hspace*{0.1cm}\includegraphics[width=0.22\textwidth]{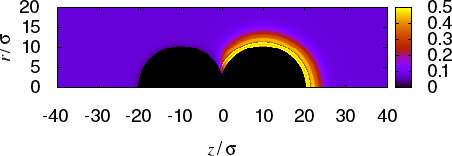}
\includegraphics[width=0.22\textwidth]{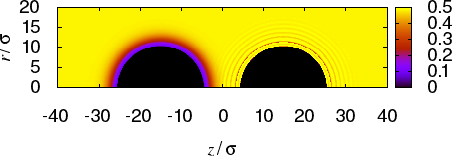}\hspace*{0.1cm}\includegraphics[width=0.22\textwidth]{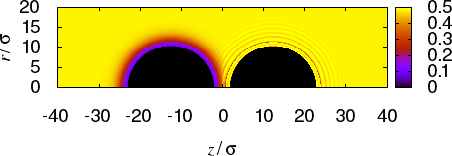}\hspace*{0.1cm}\includegraphics[width=0.22\textwidth]{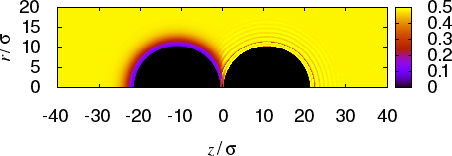}\hspace*{0.1cm}\includegraphics[width=0.22\textwidth]{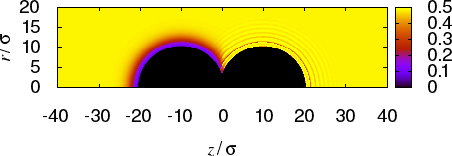}
\caption{Two-dimensional density profiles of a system consisting of a HS (left) and a SW (right) particle of (hard core) diameter $\sigma_b=20\,\sigma$ immersed in a
saturated gas (top row) and a saturated liquid (bottom row) of the SW fluid at a temperature $T/T_c=0.89$. The distances between the hard-core surfaces of the
particles are (from left to right): $h/\sigma=10, 5, 2$ and $0$.} \label{fig6}
\end{figure}

\begin{figure}
\includegraphics{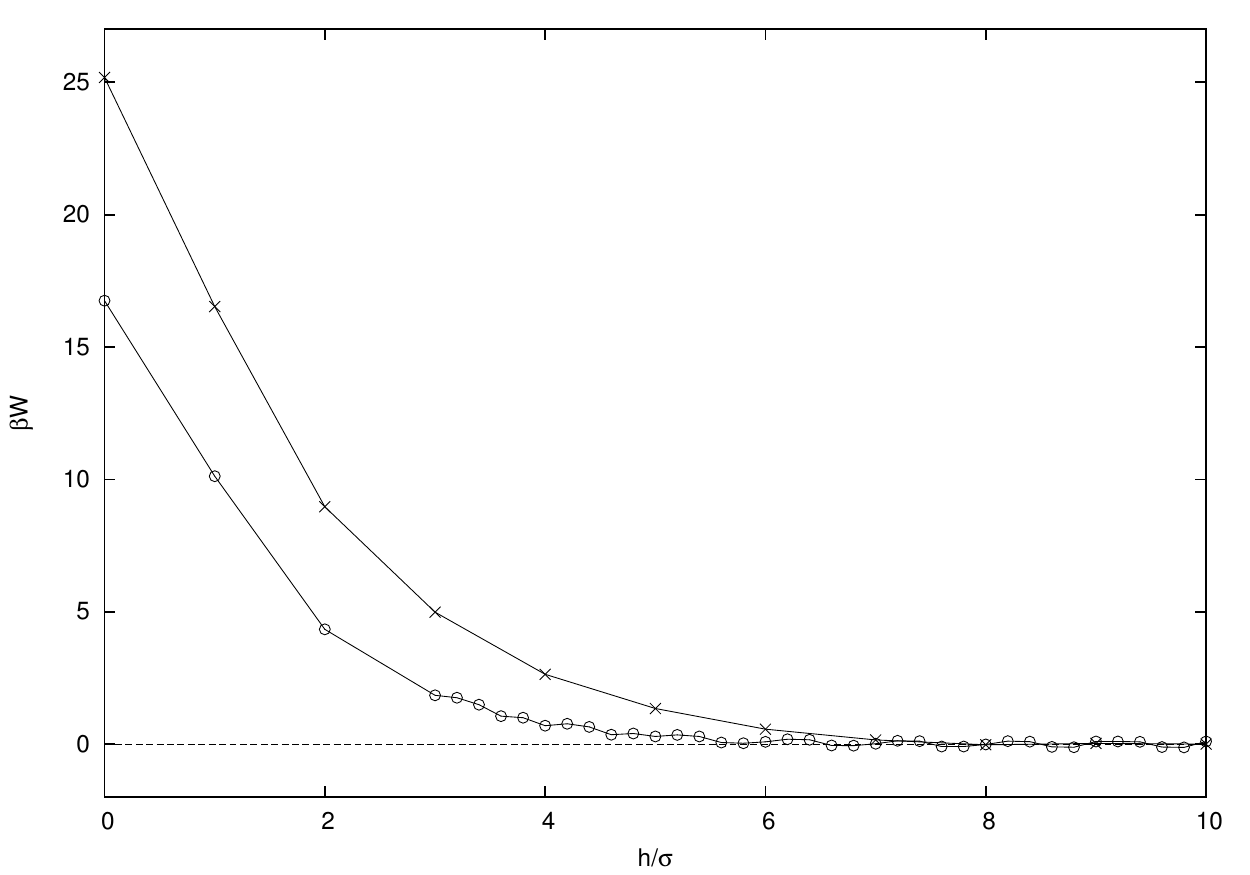}
\caption{Effective potential between a HS and a SW particle of (hard core) diameter $\sigma_b=20\,\sigma$ as a function of the distance between the the hard-core
surfaces. The pair is immersed in a saturated gas (line with crosses)  and saturated liquid (line with open circles) of the SW fluid at a temperature $T/T_c=0.89$.}
\label{fig7}
\end{figure}

We next consider the case of two ``competing'' particles that favour opposite phases. We describe this situation within Model 2, where a pair of nanoparticles is
represented  by a HS and  a SW particle. In Fig.~6 (upper row), we display several representative density profiles of different particles separations. The upper row
of Fig.~6 corresponds to a saturated vapour of the bulk SW fluid. In this case, the SW nanoparticle, if sufficiently far from the hard-sphere nanoparticle, is
surrounded by a wetting layer of thickness $\ell$. As the two particles are brought closer to each other, the wetting film is interrupted by the excluded volume of
the hard sphere. At a sufficiently small separation between the two nanoparticles, the liquid-vapour interface of the wetting film becomes the subject of two
competing forces: while the hard-sphere particle, which prefers to remain dry, pushes the interface toward the SW particle, the SW particle strives to restore the
location of the interface to a distance $\ell$ from its surface. This competition is less dramatic than in the case of a confined fluid between two parallel plates,
which would lead to the delocalisation of the liquid-vapour interface \cite{comp_walls, stewart}. As for our model with short-range interactions the thickness of the
wetting layer increases with the radius of the adsorbate rather slowly, $\ell\sim \ln R$ as $R\to \infty$, the equilibrium wetting thickness is of the order of
several solvent diameter units and so is the range of the effective force between the particles due to this mechanism. If the previous arguments are reversed and the
liquid-vapour interface is thought fixed, it becomes clear that both nanoparticles are repelled from the interface and hence from each other. As a result, the
effective potential is strongly repulsive over the range of $\ell$, as illustrated in Fig.~7.

If inserted into the saturated liquid phase of the solvent, the drying film adsorbed at the HS becomes disturbed by the presence of the SW particle that endeavours to
remain wet, as shown in Fig.~6 (lower row). This, again, induces a repulsive force between the nanoparticles over the range of the drying film thickness. The
situation is thus largely analogous to the previous case where the particles were dissolved in the gas but there are some subtle differences. Firstly, the magnitude
and the range of the effective potential for the compound dissolved in the liquid is smaller. This can be explained as follows:

Consider a system of one HS particle of radius $R$ dissolved in the saturated liquid of the SW fluid. Using a standard coarse-grained treatment, the interfacial
potential can be expressed as an excess grand potential per area of the hard sphere (see, e.g., Ref.\cite{hend}):
 \bb
 \Omega_{\rm ex}^{(\rm hs)}(\ell)/(4\pi R^2)=\gamma_{wg}(R)+\gamma_{lg}(R+\ell)\left(1+\frac{2\ell}{R}\right)+\omega(\ell)\,, \label{om_hs}
 \ee
where $\gamma_{wg}(R)$ and $\gamma_{lg}(R+\ell)$ are the surface tensions between the hard sphere and the gas and between the gas and liquid, respectively, for a
given radius of curvature of the interface. The last term in Eq.~(\ref{om_hs}) represents interaction between the liquid-gas interface and the particle-gas interface;
for our model where all the molecular interactions are short-ranged, $\omega(\ell)=a(T)\exp\left(-\frac{\ell}{\xi_g}\right)$, where $a(T)>0$ and $\xi_g$ is the bulk
correlation length of the gas. For a large SW particle dissolved in the saturated vapour of the SW fluid, the interfacial potential reads:
 \bb
 \Omega_{\rm ex}^{(\rm sw)}(\ell)/(4\pi R^2)=\gamma_{wl}(R)+\gamma_{lg}(R+\ell)\left(1+\frac{2\ell}{R}\right)+\tilde{\omega}(\ell)\,, \label{om_sw}
 \ee
where $\tilde{\omega}(\ell)=\tilde{a}(T)\exp\left(-\frac{\ell}{\xi_l}\right)$ with $\tilde{a}>0$ and where $\xi_l$ is the bulk correlation length of the liquid.

 Going back to Model 2, we can now compare the influence of the second particle on the interfacial potentials given above. To this end, we estimate how the
disruption of the fluid structure due to the presence of the second particle affects the interfacial potential pertinent to a single particle in both cases. Assuming
that the thicknesses of the equilibrium wetting and drying layers are similar (as can be inferred from Fig.~6), the two interfacial potentials given by
Eq.~(\ref{om_hs}) and Eq.~(\ref{om_sw}) differ in only the first and the third terms. At the distance of $h\approx\ell$, the repulsion between the pair of particles
will be dominated by the final terms, the asymptotic decay of which are controlled by the correlation length of the adsorbed phase. Assuming $\xi_g\approx\sigma$, the
repulsion in the liquid becomes vanishingly small for $h\gtrsim 5\,\sigma$, in accordance with Fig.~7. The particles are correlated to a larger distance in the liquid
than in the gas, i.e., $\xi_l>\xi_g$, and thus $\tilde{\omega}(\ell)$ can be expected to be slightly longer-ranged than $\omega(\ell)$. This, in turn, imply that
$W(h)$ for Model 2 is longer-ranged when the nanoparticles are dissolved in the gas.

For small separations $h<\sigma$, the dominant contribution to the effective force is due to the surface interactions between the adsorbed molecules and the
adsorbent. The work required to remove the surface particles from the region between the nanoparticles is clearly considerably higher for the high-density molecules
attracted to the SW particle than for the gas-like fluid interacting with its adsorbent only repulsively. This explains why $W(h)$ is larger for small $h$ when the
bulk fluid is gas.

The second difference between the two cases can be inferred from Fig.~6 by comparison of the fluid structure in the vicinity of the non-adsorbing particle. In the
liquid state, the additional attraction of the SW particle gives rise to a rich, oscillating structure which in turn leads to mild oscillations of $W(h)$ for large
$h$ in contrast to a monotonic decay of $W(h)$ if the bulk is vapour. The asymptotic behaviour of $W(h)$ indicates that the effective interactions can also be weakly
attractive at sufficiently large distances. The damped oscillatory structure of $W(h)$ suggests that the liquid state lies above the Fisher-Widom line \cite{fw},
while the gas state, for which the structure exhibits monotonic decay, lies below.

\section{Concluding remarks}

In this paper, we have employed non-local density functional theory to study the effective interactions between a pair of nanoparticles in a solvent exhibiting
liquid-vapour coexistence. The pair of nanoparticles plays a role of an external, axially-symmetric field, and we determined equilibrium two-dimensional density
profiles of the solvent fluid modelled grand-canonically by a SW potential. Within Model 1, the nanoparticles are identical large hard spheres. If immersed in the
vapour phase, the effective interaction corresponds to a short-range depletion potential induced by the excluded volume effects of the hard spheres. However, when
inserted into the reservoir of the saturated liquid, drying films form around the particles and mediate the effective interaction between the particles over an
arbitrarily long distance determined by the HS curvature. A strong attraction arises when the two films connect to form a bridge between the nanoparticles. The main
results of this bridging process as obtained by our microscopic DFT can be summarised as follows:

\begin{itemize}

\item The magnitude of the bridging potential found in this work was on the order of $10^2\,k_bT$. This is one order of magnitude less than the value found
in Ref. \cite{archer} and two orders less than that found in Ref. \cite{hopkins}. This reveals that the strength of the bridging interactions is  strongly
model-dependent. We attribute these differences to the disparate number of solvent particles that contribute to the bridging bond. Ref. \cite{archer} and Ref.
\cite{hopkins} address effective solvent interactions for polymers and charged particles, respectively. This means that the considered potentials are soft and the
particles may overlap. This is of course not the case for simple fluid models such as the square-well potential considered in this work. The hard core interaction
between the particles obviously reduces the number of solvent particles forming the bridge compared to soft core models. Moreover, the bridging formation considered
here was due to a connection of low density films, which strongly contrasts with the high density bridges of Refs. \cite{archer} and \cite{hopkins}. Note that the
simple mean field analysis used in Refs. \cite{archer} and \cite{hopkins} is justified for only these high density states.

\item  Supported by theoretical studies for models with short-range \cite{archer, hopkins}
and long-range \cite{yeomans, yeomans2, bauer} interactions, the bridging is generally considered to be a first-order transition. From a macroscopic viewpoint, this
picture is very reasonable because the symmetries of the bridged and unbridged states are different. However, our microscopic study reveals that the bridged state may
also develop continuously. In this case, the bridging bond forms as the wetting/drying films meet.  We have found that it is so for only relatively high radii of the
large spheres ($\sigma_b\approx20\,\sigma$) that the bridging transition becomes first-order; here, the bridging occurs before the films meet. A discussion on the
order of the transition may appear somewhat academic in view of the fact that an inclusion of fluctuations would necessarily round the transition anyway \cite{note}.
However, the study of bridging between a pair of particles is only a first step to understanding phenomena such as the flocculation in colloidal suspensions, where a
bridging connection between a pair of particles plays a role of only a single constituent in a percolating network. The entire collective (flocculation) transition
can then be genuinely first-order, provided that the bridging bond is sufficiently strong.

\item Based on the fact that the bridging transition is first order (on a mean-field level) for only large radii of the particles, one may identify  the bridging
transition as just a reminiscence of capillary condensation. In fact, for macroscopically large particles one can, in the spirit of Derjaguin's approximation, think
of the interaction between the nanoparticles in terms of the interaction between a pair of two plates. This picture becomes less satisfactory as the curvature of the
particles increases. Ultimately, the capillary condensation disappears completely below some critical value $R=R_c$, which we have identified to lay in the interval
of $(15\,\sigma,20\,\sigma)$. For radii smaller than the critical radius (corresponding to the given temperature), the bridge can still be formed \cite{note2} as the
two wetting layers merge, but the process is no longer accompanied by any phase transition. The role of the critical radius $R_c$ in bridging is thus analogous to the
role of the capillary critical distance $L_c(T)$ for capillary condensation between plates.


\end{itemize}

We have further investigated the effective interaction of two different (competing) nanoparticles using Model 2 represented by a pair of a HS and a SW particle.
Dissolved in a saturated vapour or liquid, only one wetting layer forms around a respective particle. As the particles are brought together to the distance of the
thickness of the wetting layer, the interaction between the liquid-vapour interface of the layer and the other particle leads to a strongly repulsive effective force
between the nanoparticles. Owing to the dissimilarity of the liquid and gas bulk correlation lengths, the repulsive potential is slightly longer-ranged when the
reservoir is a vapour. In this case, the magnitude of the potential is also distinctly higher than for the liquid bulk phase near a zero separation of the
nanoparticles in view of the strong interaction between the solvent molecules and the SW particle. In the gas phase, the effective potential decays monotonically,
while in the liquid phase the potential exhibits damped oscillatory structure at large distances.

In summary, we have considered two possible mechanisms that induce an effective interaction between a pair of nanoparticles mediated by a fluid solvent, a depletion
potential due to the excluded volume of the region near the contact of the nanoparticles and the effective potential induced by the presence of wetting films. The
latter is both longer-ranged and stronger than the depletion potential but of course less universal, as it requires that the solvent be near its phase separation. The
potential is strongly attractive when the nanoparticles are identical and are covered by wetting layers at separations where the wetting layers merge and form a
bridge between the nanoparticles. The bridging gives rise to a first-order phase transition if the radii of the particles are larger than a certain critical value
$R_c$. If the adsorption preferences of the nanoparticles are different and the wetting layer forms around only one nanoparticle, the effective interaction is
strongly repulsive. The third possibility that could induce the effective interaction that we have not discussed here is the critical Casimir effect. The required
thermodynamic conditions are even more restrictive in this case (the critical point of the solvent) but a qualitative picture of this interaction would be most likely
similar to the previous case, such that the interaction is attractive for identical and repulsive for opposite adsorption preferences of the nanoparticles
\cite{casimir, okamoto}.

Although we have tried to provide a rather complete microscopic description of the effective interactions near a two phase coexistence by examining four different
model situations, our results instigate further interesting questions. Most pertinently, one can ask what is behind the rather large value of the critical radius
$R_c$ and how to predict the separation at which the bridging transition occurs. To answer the latter, one needs to find a generalisation of Kelvin's equation for
spherical bodies. One can also inquire how the results would change for other molecular models, particularly for those including long-range interactions.  The
analysis of the bridging transitions can also be extended by considering two identical particles exerting attractive potentials, which would allow to address the role
of the temperature in the light of wetting transition. Although, e.g., the occurrence of the mean-field pre-wetting transition that does not occur in our Model 1
would make the problem somewhat more complex, in the main, we do not expect qualitatively different behaviour of such a model, since, as we have demonstrated, the
bridging transition is a capillary condensation pertinent to spherical walls. Finally, it would also be interesting to examine to what extent the shape of the bridge
given by the microscopic DFT matches with the macroscopic requirements of the minimal surface (catenoid) of the bridge. These and other related problems are subjects
of the current study and will be published elsewhere.


\begin{acknowledgments}

\hspace*{0.01cm}

\noindent Dedicated to the so-called Czech school of the statistical mechanics of liquids. In particular, I express my gratitude to my father, who was the first to
introduce me to the fascinating field of the liquid matter theory. I also acknowledge the financial support from the Czech Science Foundation, Grant No. 13-02938S.
\end{acknowledgments}


\begin{thebibliography}{99}
\bibitem{note}
In Ref. \cite{hopkins} it is argued that the transition would be roundend beyond the mean field theory because of the finite number of solvent particles forming the
bridge. We note that the transition would be roundend even if the nanoparticles are infinitely long cylinders, for instance, in which case the number of particles
contributing to the bridge bond is infinite because of insufficient number of dimensions along which the correlation length may diverge.
\bibitem{note2}
Strictly speaking, for small radii one should analyze a stability of the corresponding bridging films.
\end{thebibliography}
\end{document}